\documentstyle[12pt,aaspp4]{article}

\begin{document}

\title{The Small Scale Velocity Dispersion of Galaxies: A Comparison of 
Cosmological Simulations}
\author{Rachel S. Somerville\altaffilmark{1}, Joel
Primack\altaffilmark{1}, Richard Nolthenius\altaffilmark{2}}
\altaffiltext{1}{Physics Department, University of California, Santa Cruz,
CA 95064}
\altaffiltext{2}{UCO/Lick Observatory, University of California, Santa Cruz,
CA 95064}

\begin{abstract}
The velocity dispersion of galaxies on small scales ($r\sim1h^{-1}$
Mpc), $\sigma_{12}(r)$, can be estimated from the anisotropy of the
galaxy-galaxy correlation function in redshift space (Davis \& Peebles
1983). We apply this technique to ``mock-catalogs'' extracted from
N-body simulations of several different variants of Cold Dark Matter
dominated cosmological models, including models with Cold plus Hot
Dark Matter, to obtain results which may be consistently compared to
similar results from observations. We find a large variation in the
value of $\sigma_{12}(1 h^{-1} Mpc)$ in different regions of the same
simulation.  We investigate the effects of removing clusters from the
simulations using an automated cluster-removing routine, and find that
this reduces the sky-variance but also reduces the discrimination
between models. However, studying $\sigma_{12}$ as clusters with
different internal velocity dispersions are removed leads to
interesting information about the amount of power on cluster and
subcluster scales. We compute the pairwise velocity dispersion
directly in order to check the Davis-Peebles method, and find
agreement of better than 20\% in all the models studied.  We also
calculate the mean streaming velocity and the pairwise peculiar
velocity distribution in the simulations and compare with the models
used in the Davis-Peebles method. We find that the model for the mean
streaming velocity may be a substantial source of error in the
calculation of $\sigma_{12}$.

\end{abstract}


\section{Introduction}

The velocity dispersion of galaxies on small scales ($r\sim 1 h^{-1}$
Mpc), combined with cluster abundance data on intermediate scales and
the COBE normalization and galaxy peculiar velocity information on
large scales, provides a strong constraint on cosmological models by
constraining the shape of the matter power spectrum. In this paper we
investigate a method introduced by Peebles (1976; 1980) and Davis \&
Peebles (1983, hereafter DP83), which uses the anisotropy of the
redshift-space correlation function to determine the pairwise velocity
dispersion on small scales. We shall refer to this method as the
Davis-Peebles method.
 
The galaxy-galaxy correlation function  $\xi(r)$ is one of the
canonical statistics used in studying large scale structure. A related
statistic is the redshift-space correlation function, $\xi(r_{p},
\pi)$, which is a function of the components of the separation in
redshift space perpendicular ($r_{p}$) and parallel ($\pi$) to the
line of sight. If the correlation function is isotropic in real space,
it will be anisotropic in redshift space due to the peculiar
velocities of the galaxies. Hence the degree of anisotropy of
$\xi(r_{p}, \pi)$ is a measure of the moments of the peculiar velocity
distribution.

The first moment of the pairwise velocity distribution,
$\overline{v_{12}}(r)$, is proportional to $\Omega_{0}^{0.6}$ if
galaxies trace mass and density fluctuations are in the linear regime,
where $\Omega_{0}$ is the density of matter in units of the critical
density at the present epoch.  The second moment, $\sigma_{12}$, is
the velocity dispersion and measures the kinetic energy of the galaxy
distribution. This quantity has been used in combination with the
Cosmic Virial Theorem to estimate $\Omega_{0}$ (DP83), although
recently some authors have presented arguments that this calculation
is not only plagued by extremely large uncertainties from cosmic
variance (Fisher et al. 1994b), but that some fundamental assumptions
in the usual formulation of the Cosmic Virial Theorem may also be
incorrect (Bartlett \& Blanchard 1995).

The first calculation of $\sigma_{12}$ on a relatively large survey
was done by Davis and Peebles (DP83). They calculated $\sigma_{12}$
for the CfA1 redshift survey, a survey containing 1840 redshifts
covering 1.83 steradians in the North galactic hemisphere (Huchra et
al.  1983). In another paper (Somerville, Davis, \& Primack 1996,
hereafter SDP), we present a reanalysis of their work, in which we
show that $\sigma_{12}$ for this sample is extremely sensitive to the
way in which corrections for infall into the Virgo cluster are
applied, and that the value of $\sigma_{12}$ for this survey is
dominated by the clusters. The same calculation was done on the
Southern Sky Redshift Survey(SSRS1) (da Costa et al. 1991), with
results of $\sigma_{12}(1 h^{-1} Mpc) \equiv \sigma_{12}(1) \sim 300$
km/s (SDP; Davis 1988). Recently $\sigma_{12}$ has also been
calculated for the IRAS survey (Fisher et al. 1994b), the CfA2/SSRS2
survey (Marzke et al. 1995), and the Perseus-Pisces survey (Guzzo et
al. 1995). These calculations have shown a range of values of
$\sigma_{12}(1)$ from 272 km/s for SSRS2 to 769 km/s for
Perseus-Pisces (see Table 1 of SDP for a summary).

The values of $\sigma_{12}$ usually quoted for simulations are
calculated by measuring the dispersion of the pairwise peculiar
velocity field directly using the full three-dimension position and
velocity information for the halos (Davis et al. 1985; Gelb \&
Bertschinger 1994; Klypin et al. 1993). In real redshift surveys, not
only are there errors introduced by edge effects and selection
effects, but $\sigma_{12}$ is extracted by fitting a model to the
correlation function in redshift space, $\xi(r_{p}, \pi)$.  This
quantity is quite noisy especially for samples with small numbers of
galaxies. In addition the procedure involves a number of
assumptions. It is a reasonable question, therefore, whether the
values from simulations may be meaningfully compared to the
observational values. Zurek et al. (1994) applied the Davis-Peebles
method to mock redshift surveys extracted from simulations of a
standard Cold Dark Matter (CDM) model, and found that this yields a
rather large range of values for $\sigma_{12}$. So far there has not
been a comparison of $\sigma_{12}$ calculated using the Davis-Peebles
method on ``observed'' simulations of different cosmological
models. In this paper, we investigate the robustness of $\sigma_{12}$
using mock redshift surveys extracted from several different cold dark
matter dominated models, and the ability of this statistic to
discriminate between such models. We estimate the sky-variance and
cosmic variance of the statistic and identify sources of error in the
Davis-Peebles method. We also investigate the effects of removing
clusters from the samples.

\section{Cosmological Models and Simulations}
\label{sec-sims}
The models and simulations are described in detail in Klypin,
Nolthenius \& Primack (1995, henceforth KNP).  All of the models
studied here have Gaussian initial fluctuations with a
Harrison-Zel'dovich scale invariant spectrum, density parameter
$\Omega=1$ and Hubble parameter $h = 0.5$ ($H_0 = 100h \: \rm{km
s^{-1} Mpc^{-1}}$).

The models represented are two variants of Cold Dark Matter (CDM) and
one variant of Cold plus Hot Dark Matter (CHDM). The ``standard'' Cold
Dark Matter model assumes that the fraction of the mass in the
universe made of baryons is $\Omega_{b} = 0.1$ with the rest of the
mass made up of a species of non-relativistic, dissipationless
particle (the cold dark matter).  There are different ways of
normalizing the spectrum --- ``unbiased'' (b=1), which assumes that
the galaxy fluctuations trace the dark matter fluctuations, and
``biased'', which assumes that perhaps only especially large amplitude
fluctuations lead to galaxy formation, i.e.
$\left(\frac{\delta\rho}{\rho}
\right)_{galaxy} = b \: \left(\frac{\delta\rho}{\rho}\right)_{mass}$. We
have analyzed an ``unbiased'' (b=1) simulation (which we call CDM1)
and a b=1.5 simulation (CDM1.5).  In linear theory, this corresponds
to rms fluctuations of mass in a sphere of $8h^{-1}$ Mpc radius
$\sigma_8=1$ for the unbiased model, $\sigma_8=0.667$ for the biased
model, or a quadrupole in the angular fluctuations of the cosmic
microwave background of $Q_{ps-norm}=12.8 \mu K$ or $Q_{ps-norm}=8.5
\mu K$ respectively (KNP). The value of $Q_{ps-norm}$ for the unbiased model
was consistent with the lowest normalization quoted in the first-year
COBE results (Smoot et al. 1992). It is too low in view of the more
recent COBE results, which give $Q_{ps-norm} = 18 \mu K$ (Gorski et
al. 1996), but this model is still of considerable interest. The
biased model (CDM1.5) is not compatible with the COBE measurement of
$Q_{ps-norm}$; however, it does predict approximately the correct
number density of clusters (Frenk et al. 1990), and is useful for
comparison.

The Cold plus Hot Dark Matter (CHDM) model assumes that the
non-baryonic matter is made of a cold component, as before, and a hot
component, generally thought to be a massive neutrino. In ``standard''
CHDM, the ratio of cold to hot to baryonic matter is
$\Omega_{c}/\Omega_{\nu}/\Omega_{b} = 0.6/0.3/0.1$ , corresponding to
a single massive neutrino with a mass of about 7 eV.  The power
spectrum in our model is normalized to $\sigma_8 = 0.667$,
corresponding to $b=1.5$ or $Q_{ps-norm} = 17 \mu K$. More recent data
and analyses favoring a smaller fraction of the mass in neutrinos,
$\Omega_{\nu} \simeq 0.2$, are summarized in Primack et al. (1995) and
Liddle et al. (1996). Our preliminary analysis of new simulations with
$\Omega_{\nu} = 0.2$ indicate that $\sigma_{12}$ in this model does
not differ substantially from the $\Omega_{\nu} = 0.3$ model on the
scales of interest in this paper.

The simulations were done using a standard Particle-Mesh (PM) code
with a $512^3$ force mesh. Each simulation has $256^3$ cold particles,
and the CHDM simulations have two additional sets of $256^3$ hot
particles with random thermal velocities corresponding to Fermi-Dirac
distribution (KNP). The size of the computational box is 100 Mpc
(i.e. $50 h^{-1} Mpc$ for $h=0.5$) and the smallest resolved co-moving
scale is $\sim 100 h^{-1}$ kpc.

The initial fluctuations were generated with the same random numbers
for CHDM1 and both CDM simulations. After running the CHDM1
simulation, it was discovered that there were two mistakes in the
initial conditions (see KNP), which fortunately are in phase and
largely cancel. A new simulation with the same cosmological parameters
but different initial conditions (and mistakes corrected) was run ---
we call this CHDM2. Extensive comparison of the two simulations has
shown that the power and velocity differences on small scales are no
more than 5\% at $z<7$. In addition, there was a statistical fluke (of
about 10\% probability) in the initial conditions for both CDM
simulations and CHDM1: the amplitude of the longest waves was a factor
of 1.3-1.4 larger than that expected for the ensemble average, so the
power on large scales is approximately a factor of two larger than
typical. However, this could be considered a compensation for the
finite size of the box, or in the case of CDM1, as compensation for
the low normalization compared to COBE. Also, comparing CfA1 to the
much larger APM survey shows that the CfA1 region has unusually high
power at these scales (Nolthenius, Klypin \& Primack 1996 (NKP), Baugh
\& Efstathiou 1993). In any case comparison of CHDM1 and CHDM2, which
has a typical amount of power, gives us a measure of cosmic variance.
 
In order to compare our simulations with observations, we need to
identify objects which correspond to galaxies. There is no
satisfactory proscription to do this in simulations containing only
dissipationless dark matter (Summers, Davis, \& Evrard 1995). One
approach is to assume that galaxies form in regions where the dark
matter has collapsed to a sufficiently high density.  However, a
well-known problem with this procedure is that when the dark matter
halos merge, they quickly lose all discernable substructure. This
results in a halo mass distribution which includes a few very large
mass ($\sim10^{13}-10^{15} M_{\odot}$) halos in the final epoch of the
simulations which should probably be associated with the dark matter
in cluster cores rather than with the halos surrounding single
galaxies (Katz \& White 1993; Gelb \& Bertschinger 1994).  In
simulations with hydrodynamics and gas, the baryons can lose energy
and form smaller clumps which remain distinct within the overmerged
dark matter halos (c.f. Evrard, Summers, \& Davis 1994; Hernquist \&
Katz 1989; Cen \& Ostriker 1992).  However, associating these baryon
clumps with galaxies is still an oversimplification as the details of
galaxy formation depend on many interdependent processes including
dissipation, gas dynamics, star formation, and energy feedback from
supernova. Exactly how these processes affect galaxy formation is not
well understood and it is not possible to include these effects in
N-body simulations on cosmological scales with present computing
capabilities. Clearly, for the present study of large scale structure,
it is necessary to attempt to make use of information from
dissipationless simulations as best we can. Therefore, we have adopted
a scheme, informed by results from simulations with gas and
hydrodynamics (Evrard et al. 1994), for ``breaking-up'' the overmerged
dark matter halos. Although this procedure is ad hoc and has many
limitations, it does go a step further towards enabling a realistic
comparison between real redshift surveys and simulations than many
previous analyses of simulations.

The procedure for breaking up the halos, assigning luminosities, and
forming magnitude limited ``sky-catalogs'' which mimic the CfA1
redshift survey is described in detail in NKP.  First, halos were
identified as mesh cells with a sufficiently high dark particle mass
overdensity $\delta\rho/\rho$ at the end of the simulation ($z=0$).
Halos with a mass above a certain cutoff were broken up and the
fragments were assigned masses according to a Schechter
distribution. Fragment velocities were chosen randomly from a Gaussian
distribution using the rms velocity of the nearest neighbors as the
dispersion. Luminosities were chosen randomly from a Schechter
distribution with the same parameters as the CfA1 catalog and assigned
to the halo fragments assuming that higher luminosity corresponded to
higher 1-cell mass. This results in a distribution with the same
selection function as the CfA1 catalog. The broken-up dark matter
halos are hereafter somewhat metaphorically referred to as
``galaxies''.

To construct the mock CfA1 catalogs, six different ``home galaxies''
were selected for each simulation, in such a way as to mimic the
conditions of the CfA1 redshift survey: the home galaxies were
required to lie in an area with local galaxy density within a factor
of 1.5 of the local CfA1 density, and to have a Virgo-sized cluster
about 20 Mpc away.  An ``observer'' was placed on each home galaxy and
radial velocities along the line of sight were calculated for each
galaxy. Catalogs were created with the same angular boundaries as the
CfA1 North survey ($b>40^{\circ}, \delta > 0^{\circ}$, 1.83 sr), and
also all-sky catalogs containing all galaxies with $|b| > 10^{\circ}$
(10.4 sr).

In order to check the effectiveness of our break-up procedure, we have
verified that the galaxy-galaxy correlation function in real space
follows a power law all the way down to the resolution limit. Without
break-up, the correlation function falls below the power law on small
scales.  We have also checked that hierarchical scaling holds for the
galaxies; i.e. that the volume-averaged three-point function is
proportional to the averaged two-point function squared. We have found
that the reduced skewness $S_{3} =
\overline{\xi}_{3}(r)/\overline{\xi}_{2}(r)^2$ is constant to a good
approximation from $r=0.5$ Mpc to $r = 20$ Mpc. These results suggest
that our break-up procedure leads to a halo distribution with the
expected clustering properties, at least to this order on these
scales.

\section{Method}
\label{sec-method}
In this section we briefly describe the method used to extract the pairwise
velocity dispersion $\sigma_{12}$ from the redshift-space correlation
function $\xi(r_{p}, \pi)$. Readers should refer to DP83 and Fisher
et al. (1994a, b) for more details of the method. 
 
The correlation function in redshift space, $\xi(r_{p},
\pi)$, is estimated by counting the number of pairs in a bin in
$r_{p}$ (separation perpendicular to the line of sight) and $\pi$
(separation parallel to the line of sight). It is normalized by
constructing a catalog of Poisson distributed points with the same
selection function and angular limits as the data, and counting pairs
between the data and the Poisson catalog:
\begin{equation}
1+\xi(r_{p}, \pi) = \frac{n_{R}}{n_{D}} \cdot \frac{DD(r_{p},
\pi)}{DR(r_{p}, \pi)} \label{eq:xi} 
\end{equation}
where DD is the number of pairs between data and data, and DR is
number of pairs between the data catalog and a Poisson catalog. The
quantities $n_{R}$ and $n_{D}$ are the minimum variance weighted
densities (see Davis \& Huchra 1982) of the Poisson and data catalogs,
respectively. In practice, we use a large ensemble of Poisson catalogs
in order to reduce shot noise and to ensure that no bin has zero pair
count.

Let $F(\bf{w \mid r})$ be the distribution function of velocity
differences $\bf{w}$ for pairs of galaxies with vector separation
$\bf{r}$, and $f(w_{3} \mid r)$ the velocity distribution function
averaged over the directions perpendicular to the line of sight.  The
first moment of $F(\bf{w \mid r})$ is $\overline{v_{12}}(r) \equiv -
\langle \bf{[v(x)-v(x+r)]
\cdot \hat{r}} \rangle$
where the average is number (not volume) weighted. This quantity is
also known as the mean streaming velocity, and assuming isotropy it is
a function only of the magnitude of $\bf{r}$. Correspondingly, the
first moment of $f(w_{3} \mid r)$ is $\langle w_{3} \rangle = y \:
\overline{v_{12}}(r)/r$ where $y$ is the component of $\bf{r}$ along the
line of sight. 

A reasonable form for the distribution function $f(w_3 \mid r)$,
parameterized by its moments, is
\begin{equation}
f(w_3 \mid r) \propto \exp \left(\nu\left[\frac{w_3(r)-\langle w_3(r) 
\rangle}{\sigma_{12}(r)}\right]^{n}\right)
\end{equation}
It has been found emperically from studying observations and N-body
simulations (Peebles 1976; Fisher et al. 1994b; Marzke et al. 1995)
that on small scales an exponential form (n=1) fits the data better
than a Gaussian (n=2) or any higher power of the argument. We have
also tested this assumption for our N-body simulations and find
excellent agreement with an exponential in all the models up to scales
of $r\sim 5 h^{-1} Mpc$.  Recently, Sheth (1996) gave a derivation of
the exponential form for the distribution function using the
Press-Schechter approach. Adopting this form for $f(w_3
\mid r)$, and using $r^{2} = r_{p}^{2} + y^{2}$ and $w_3 = \pi - y$ we
have:
\begin{equation}
1 + \xi(r_{p}, \pi) = \frac{H_{0}}{\sqrt{2}}
\int \frac{dy}{\sigma_{12}(r)} \left[1 +
\xi(r)\right] \exp \left[\frac{-\sqrt{2}}{\sigma_{12}(r)}
 \left| \pi- H_0 y \left[1 - \frac{\overline{v_{12}}(r)}{H_{0}r} 
\right] \right| \right]   \label{eq:modelxi}
\end{equation}
An approximation based on self-similar solutions of the BBGKY
hierarchy leads to a form for $\overline{v_{12}}(r)$ (Davis \& Peebles
1977):
\begin{equation}
\overline{v_{12}}(r) = \frac{FH_{0}r}{[1 + (r/r_{0})^{2}]}  \label{eq:infall}
\end{equation}
F is an adjustable parameter of the model. The assumption of stable
clustering, i.e. that the collapse of the cluster is exactly balanced
by the Hubble flow, corresponds to F = 1. The results presented here
are for F = 1 unless stated otherwise.  Later on in this
paper we investigate this model, including the validity of the stable
clustering assumption, by calculating $\overline{v_{12}}(r)$ in the
N-body simulations. The velocity dispersion $\sigma_{12}$ is then obtained
by fitting the model of equation~(\ref{eq:modelxi}) to $\xi(r_{p},
\pi)$ estimated from equation~(\ref{eq:xi}).

\section{Results}

One approach to comparing observations to simulations is to try to
translate the space of observed quantities (i.e. redshift) into the
space of the simulations (real space). Reflecting this approach, DP83
attempted to correct the measured redshifts by modeling the flow field
around Virgo, which is in the foreground of the CfA1 survey. However,
we have shown that the value of $\sigma_{12}$ depends sensitively on
the details of these corrections (SDP).  Another approach is to
``observe'' the simulations to form simulated redshift surveys, which
attempt to mimic as closely as possible what astronomers would
actually observe if they lived in the universe of that simulation, and
to then analyze the simulated and observed catalogs in exactly the
same way.  Taking this approach, we have selected catalogs from the
simulations requiring that a Virgo-sized cluster appear in the
foreground as described in section \ref{sec-sims}, and do not include
corrections for cluster infall in any part of the procedure.

Contour plots of $\xi(r_{p}, \pi)$ for the 10.4 sr sky-catalogs are
shown in Figure~\ref{fig:xicont}. For each simulation, $\xi(r_{p},
\pi)$ has been averaged over 6 views and smoothed. (For purposes of the
contour plots only, $\xi(r_p, \pi)$ was calculated with linear bins in
$r_p$ and $\pi$. For the rest of the calculation, logarithmic bins
were used in $r_p$ and linear bins in $\pi$). There is a visible
difference in the shape of the contours for the different models. In
the absence of peculiar velocities, the contours would be perfectly
circular segments. However, the velocity dispersion of clusters causes
structures to appear elongated along the line of sight for small $r_p$
(the familiar ``fingers of god''), and infall causes the contours to
appear compressed in the line of sight direction for larger
$r_p$. This effect is much more pronounced in the CDM models as
expected due to the larger velocity dispersions and infall velocities.

The values of $r_0$ and $\gamma$ obtained from the inversion of
$w(r_p)$ (see DP83) and used in the fit for $\sigma_{12}$ are given in
Table 3. The results for $\sigma_{12}(r)$ are shown in
Figure~\ref{fig:sigma}. The points are the average of the results for
the six views, and the error bars are standard deviations for the
different views. These error bars reflect the ``sky variance'', or
variation within the same simulation when viewed from different
points, which for this relatively small box size is likely to be an
underestimate of the true cosmic variance. The difference between
CHDM1 and CHDM2, about 200 km/s at $1 h^{-1}$ Mpc, may give a better
estimate of the possible cosmic variance. This is consistent with the
large variance seen in recent calculations of $\sigma_{12}$ in
different redshift surveys (see SDP).

\subsection{Removing Clusters}
\label{sec-rmclust}
We attempted to address this apparent non-robustness of $\sigma_{12}$
by developing an algorithm to automatically remove clusters from
catalogs in a way which could be applied consistently to both real and
simulated data.  It is no surprise that rich clusters are a major
source of sample-to-sample variation in $\sigma_{12}$. Because
$\xi(r_{p}, \pi)$ is a pair-weighted statistic, clusters tend to
dominate it. We saw in SDP that removing the Virgo cluster from the
CFA1 survey significantly reduced $\sigma_{12}$, consistent with the
findings of Zurek et al. (1994). The SSRS1 survey does not contain any
clusters as rich as the Virgo or Coma clusters found in CfA1, and
$\sigma_{12}$ is much smaller for SSRS1 ($\sigma_{12}(1h^{-1}Mpc)
\simeq 320$ km/s) than for CfA1 ($\sigma_{12}(1h^{-1}Mpc) \simeq 620$
km/s)(SDP). Fisher et al. (1994b) found $\sigma_{12}(1)$ for the IRAS
redshift survey to be only about 320 km/s, much lower than the
result for CfA1. The IRAS survey is dominated by dusty spiral galaxies
and undercounts cluster centers by about a factor of 2 relative to
optically selected surveys. Marzke et al. (1995) also found that
removing Abell clusters of richness class R$\geq$1 significantly
reduces $\sigma_{12}$ in the CFA2/SSRS2 survey, and that $\sigma_{12}$
changes more drastically in regions of the survey where there were
many rich clusters to begin with.  In view of this evidence, we
thought it interesting to see what happens to $\sigma_{12}$ when the
clusters are removed from the simulations. We therefore developed a
method to remove clusters which can be applied consistently to both
simulations and observations.

Our algorithm is as follows. We divide the catalog into bins and
identify bins with density fluctuations larger than a specified
cutoff. We then calculate the luminosity weighted centroid of all
points lying within a cylinder of radius $r_{c}$ and redshift interval
$2h_{c}$ centered on the bin where the fluctuation was found. We use
$h_c = 1000$ km/s, and adjust $r_c$ as described later. We then take a
new cylinder centered on the centroid, calculate the new centroid, and
continue to iterate in this way until the position of the centroid
changes by less than some small value. Finally, we calculate the
velocity dispersion of all the galaxies lying within the cylinder
around the converged centroid, and cut all the galaxies in this
cylinder only if the velocity dispersion is greater than a cutoff
$\sigma_{c}$. We have tested our algorithm by visualizing the surveys
to make sure that the regions which are cut correspond to those that
would be identified visually. The number of clusters identified
depends on both parameters $r_c$ and $\sigma_c$. We chose a value of
$r_c$ by plotting profiles of the clusters and identifying the radius
at which the number density had dropped to the background level. The
fiducial value we chose, $r_c = 2.0 h^{-1}$ Mpc, is fairly close to
the usual Abell radius. Table 1 shows the results of varying
$\sigma_c$. As $\sigma_c$ is lowered, the algorithm identifies more
and more objects as ``clusters'' --- of course as we go to lower values
of $\sigma_c$ we are really starting to identify objects which we
would normally refer to as groups. We select $\sigma_{c} = 500$ km/s
as corresponding to what are usually referred to as clusters because
at this cutoff approximately 4-9 \% of the galaxies are in clusters,
which roughly agrees with observations.  Table 2 shows properties of
clusters identified at $\sigma_{c} = 500$ km/s.

The results for $\sigma_{12}(r)$ after cluster removal with $r_c = 2.0
h^{-1}$ Mpc and $\sigma_c = 500$ km/s are shown in
Figure~\ref{fig:sigmanoclusters}. The sky-variance has decreased
somewhat for the CDM models but not significantly for the CHDM models.
The cosmic variance between the two CHDM models has decreased
slightly: $\sigma_{12}(1)$ for CHDM1 and CHDM2 now differs by about
150 km/s instead of 200 km/s. However $\sigma_{12}(1)$ for the biased
CDM model now lies between the two values for CHDM1 and CHDM2, and
even standard CDM gives a value of $\sigma_{12}(1)$ which is within
the view-to-view error bars of the CHDM1 value.  Therefore, the small
improvement in robustness acheived by removing the clusters appears to
have been attained at the expense of discrimination between models.
This problem cannot be solved by using a different value for
$\sigma_c$, as can be seen from Figure~\ref{fig:sigc_vs_sig12}. The
discrimination becomes worse as we go to lower values of $\sigma_c$,
and for higher values there is very little change in the values of
$\sigma_{12}$. The fact that $\sigma_{12}(1)$ appears to converge to
almost the same value for all of the models as we lower $\sigma_c$ is
probably an indication that, once we remove the collapsed objects,
$\sigma_{12}$ is really a measure of $\Omega_0$. In fact, originally
this statistic was not designed as a tool to discriminate between
models. It was hoped that it would be useful as a measure of
$\Omega_{0}$ (or of $\beta \equiv f(\Omega)/b$, to be precise, where b
is the bias factor defined in section \ref{sec-sims}) (DP83, Davis
1995). It would be interesting to see what would happen as we remove
clusters in simulations with different values of $\Omega_0$.

The change in $\sigma_{12}$ as a function of $\sigma_c$ has a
different shape for the different models. In
Figure~\ref{fig:ncut_vs_dsig12}, we show the change in
$\sigma_{12}(1)$ as a function of the number of clusters cut at
different values of $\sigma_c$.  It is interesting that the curve for
CHDM1 lies on top of the curve for CHDM2 on this plot even though the
values of $\sigma_{12}$ are very different. What this seems to
indicate is that although the curve for CHDM1 extends further to the
right, indicating that CHDM1 has more clusters than CHDM2 (because of
the excess large scale power), the clusters themselves have the same
velocity structure, because the small scale power is the same. The CDM
models have more power on small scales and the clusters have more
kinetic energy, as indicated by the steeper rise in
$\Delta\sigma_{12}(1)$. In addition the curves extend further to the
right than CHDM2 because the CDM models also have a larger number
density of clusters than a typical CHDM model. Plotted in this way,
this quantity may give interesting information on the amount of power
present on cluster and subcluster scales. It appears to be more robust
and discriminatory than the actual value of $\sigma_{12}(1)$. However,
it is not very useful at present because existing redshift surveys do
not contain enough clusters to allow this quantity to be calculated in
the region where it is discriminatory. It would be interesting to
study this in future large volume surveys with good sky coverage.

\subsection{The Form of the Pairwise Peculiar Velocity Distribution and
the Mean Streaming.}
\label{sec-pvd}
A form for the pairwise peculiar velocity distribution must be assumed
in the procedure we have used to calculate $\sigma_{12}$. As we
discussed briefly, the form used by DP83 and other authors who have
recently completed similar analyses is an isotropic exponential, which
was proposed by Peebles (1976) because it appeared to fit
observations. This functional form was further investigated by Fisher
et al. (1994b) and Marzke et al. (1995) and found to be consistent
with the IRAS and CFA2/SSRS2 redshift surveys, and with Cold Dark
Matter N-body simulations.  However, we thought it would be
interesting to study the form of the pairwise velocity distribution in
real space for our different models to re-evaluate whether this is the
most appropriate form. We also take advantage of the fact that unlike
in real observations, in the simulations we know the real-space
positions of the galaxies and their peculiar velocities separately. We
compare the ``true'' value of $\sigma_{12}$ obtained by fitting
directly to the peculiar velocity distribution with the results of the
Davis-Peebles procedure. We may then evaluate whether the ``true''
value of $\sigma_{12}$, uncomplicated by extracting it from the
redshift space data, is a robust discriminator between models. In this
way we can test the accuracy of the Davis-Peebles method for
extracting $\sigma_{12}(r)$.

We find that the exponential is an excellent fit to the distribution
on small scales in all the models.  On scales of $r \sim 5 h^{-1}$ Mpc
the distribution begins to look a bit flatter at small $w_3$ than the
exponential, and the distribution begins to approach a Gaussian at
even larger separations. However, at the scales where this procedure
is used ($\sim 1 h^{-1}$ Mpc), using this model is not likely to be a
significant source of error.

As we mentioned in section~\ref{sec-method}, following DP83 (and
subsequent workers) we have modeled the mean streaming velocity
$\overline{v_{12}}(r)$ by equation~(\ref{eq:infall}) in our fitting
procedure. Marzke et al. (1995) and others have suggested that the
observed scale dependence of $\sigma_{12}$ may be merely an artifact
of this term. We have investigated this quantity by computing it
directly in the simulations. It is shown in Figure~\ref{fig:v12} along
with the model for F=1 and F=0.5. The general form of the model holds
on intermediate scales, but there is a large amount of variation
between the different views, and the prediction of the model on scales
of $1 h^{-1}$ Mpc is in many cases quite inaccurate. The CDM models
show a rather large negative mean streaming velocity on small scales,
which may correspond to shell crossing. On average the model with
$F=1$ appears to overpredict $\overline{v_{12}}$ for CHDM, while
$F=0.5$ may be a better fit. This is surprising because $F=1$
corresponds to streaming which exactly cancels the Hubble flow on
scales less than the correlation length (stable clustering), $F>1$
corresponds to collapse on those scales, and $F<1$ indicates that the
clusters are not collapsing, and in fact are expanding. This implies
that stable clustering on is not a good assumption for the CHDM
models.

To show the dependence of the results on the assumed value of $F$, in
Figure~\ref{fig:signoinfall} we show the results obtained for
$\sigma_{12}$ (using the Davis-Peebles method on the ``observed''
sky-catalogs, as before) when we do not include the streaming model in
the fit, i.e. when we set $F=0$. This systematically reduces the
values of $\sigma_{12}$ that we obtain, and changes the scale
dependence as expected. The CHDM models are now entirely flat over
$r_p$. It is also possible to allow $F$ to be fit as a free
parameter. The problem with this approach is that there is a
degeneracy between $F$ and $\sigma_{12}$ --- we tended to obtain lower
values of both parameters when we allowed $F$ to be fit freely. In
addition, having two free parameters increases the error on the
fit. Apparently, the use of this model for the mean streaming,
especially with the assumption of stable clustering ($F=1$) could be a
substantial source of error in the procedure.

Figure~\ref{fig:truesigma} shows the results obtained when
$\sigma_{12}$ was fit directly to $f(w_3|r)$ using the peculiar
velocity information from the simulations. Although we use the real
space positions and peculiar velocities of the galaxies, we use the
same galaxies as those in the magnitude-limited sky-catalogs in order
to facilitate a direct comparison of $\sigma_{true}$ with the value
obtained by the Davis-Peebles method ($\sigma_{DP}$). Table 3 shows
$\sigma_{DP}$ with and without the streaming model and $\sigma_{true}$
at $1 h^{-1}$ Mpc, averaged over the six views. The results for
$\sigma_{DP}$ are within 20\% of $\sigma_{true}$ for all the
simulations. The agreement with $\sigma_{true}$ is better for all the
simulations except CHDM2 if we do not include the streaming model
($F=0$).  The view-to-view variation is comparable for $\sigma_{DP}$
and $\sigma_{true}$, which suggests that the variation is not noise
introduced by the Davis-Peebles method of extracting $\sigma_{12}$ but
that it is intrinsic in the simulations and presumably in the real
Universe.

\section{Discussion and Conclusions}

For the N-body simulations of three different cosmological models that
we have analyzed, we find that the values of $\sigma_{12}$ are
considerably higher for CDM models than for CHDM models. The CHDM
models, which give $\sigma_{12}(1) \sim 540 (440)$ km/s for CHDM2 to
740 (600) km/s for CHDM1 are perhaps more consistent with the body of
observational values taken as a whole. (Numbers in paranthesis are for
$F=0$). Although $\sigma_{12}(1) = 647$ km/s for CfA2 North is
marginally consistent even with unbiased CDM ($\sigma_{12}(1)
\sim 1024 (880)$ km/s), $\sigma_{12}(1)$ for CfA2 South (367 km/s) and
SSRS2 (272 km/s) are not (Marzke et al. 1995).  However, because of
the problems with galaxy identification discussed in Section
\ref{sec-sims}, we do not think that any model studied here should be
ruled out on the basis of these results.  Any existing large volume
N-body simulations would have similar problems. The point of this
paper is precisely to suggest that it is premature to use
$\sigma_{12}$ to draw any strong conclusions about cosmological
models.  However, studying the simulations has given us other
interesting information.

We have estimated the expected sky-variance and cosmic variance of
$\sigma_{12}$ in our models. The following values are quoted for
$\sigma_{12}(1 h^{-1}Mpc)$. The sky-variance (calculated as the
standard deviation over six mock catalogs) ranges from $\sim 40$ km/s
for CHDM2 to $\sim 145$ km/s for CDM1. The cosmic variance (between
two simulations with different initial conditions) for CHDM is $\sim
200$ km/s. The errors usually quoted for $\sigma_{12}$ are formal
errors on the fit (for which we obtain typically $\sim 40$ km/s) and
in general are underestimates of the actual statistical errors.  We
evaluate the accuracy of the Davis-Peebles method for extracting
$\sigma_{12}$ from redshift catalogs by comparing with the results of
computing the velocity dispersion directly in our mock catalogs. We
obtain agreement of better than 20\% for all of our models. This leads
us to interpret the large range of values of $\sigma_{12}$ obtained
from different redshift surveys as an intrinsic variation due to the
sensitivity of the statistic to the clusters contained in the sample,
rather than being due to errors in the method.

We have investigated the effects of removing clusters from the samples
using an automated procedure. It was hoped that this might make
$\sigma_{12}$ a more robust statistic. However, we found that although
this reduces the sample-to-sample variation in $\sigma_{12}(r)$ by a
small amount, it actually reduces the ability of the statistic to
discriminate between the cosmological models we studied. This may be
due to the fact that all of our simulations are of $\Omega = 1$
models, and that once clusters are removed $\sigma_{12}$ is really a
measure of $\Omega_{0}$. However, our study of the simulations
suggests that the change in $\sigma_{12}(1)$ as a function of the
number of clusters removed may be an interesting quantity to study in
future redshift surveys.

We find that an exponential form for the pairwise peculiar velocity
distribution is an excellent approximation on small scales ($r<5
h^{-1}$ Mpc) in all the models studied. Measuring the mean streaming
$\overline{v_{12}}(r)$ directly from the simulations revealed that
although the general form of the model used in the Davis-Peebles
method does hold, on small scales the measured values may deviate from
it considerably. We found that stable clustering is a reasonable
approximation in the unbiased (b=1) CDM model but not in the CHDM
models. The use of the BBGKY model for the mean streaming, especially
with the assumption of stable clustering ($F=1$) could be a
substantial source of error in the Davis-Peebles method.

We have shown that $\sigma_{12}$ is very sensitive to both the number
and the properties of the clusters in a sample. This makes
$\sigma_{12}$ a poor constraint on cosmological models given the
current situation with regards to both simulations and
observations. Even the largest existing redshift surveys do not
represent a fair sample of rich clusters. Also, current N-body
simulations do not simulate clusters realistically because cluster
properties are probably sensitive to non-gravitational physics such as
gas hydrodynamics, star formation, and supernova feedback --- effects
which are impossible to include in large volume simulations with
current computing capabilities. As larger redshift surveys become
available and it becomes possible to simulate clusters in a
cosmologically relevant volume, perhaps the robustness of
$\sigma_{12}(1)$ will improve. In fact, we have suggested a way in
which the very sensitivity of $\sigma_{12}$ to the properties of
clusters could be used to define an interesting statistic for
characterizing large scale structure in larger samples. In the
meantime it is worthwhile to work on developing statistics which are
discriminatory but less sensitive to the properties of clusters.

\clearpage

\begin{center}
Acknowledgements
\end{center}

\begin{acknowledgements}
We thank M. Davis for useful discussions. RSS is supported by a GAANN
fellowship from the NSF. JRP and RN are partially supported by grants
from the NSF and NASA.
\end{acknowledgements}

\clearpage

\begin{table*}
\begin{tabular}{cccccc}
&$\sigma_{cut}$ (km/s) & $N_{cl}$ & $N_{gal}$ & $f_{cl}$ & $n_{cl}
(10^{-7} Mpc^{-3})$\\
\tableline
CDM1    & 600 & 8 & 106 & 1.2 & 1.5\\
        &500 & 37 & 773 & 8.8 & 6.9\\
        &400 & 78 & 1605 & 18 & 14\\
        &300 & 116 & 2254 & 26 & 22 \\
\tableline
CDM1.5  &600 & 13 & 204 & 2.3 & 2.4\\
        &500 & 45 & 776 & 8.7 & 8.3\\
        &400 & 81 & 1398 & 16 & 15 \\
        &300 & 121 & 2142 & 24 & 22\\
\tableline
CHDM1   &600 & 5 & 89 & 1.0 & .93\\
        &500 & 25 & 530 & 6.0 & 4.6\\
        &400 & 78 & 1738 & 20 & 14\\
        &300 & 130 & 2865 & 33 & 24 \\
\tableline
CHDM2   &600 & 0 & 0 & 0 & 0\\
        &500 & 15 & 368 & 4.3 & 2.8\\
        &400 & 34 & 800 & 9.4 & 6.3\\
        &300 & 82 & 1807 & 21 & 15\\
\tableline
\end{tabular}

\caption{Properties of clusters identified using the algorithm described in
the text, averaged over six views, for the 10.4 sr
sky-catalogs. $N_{cl}$ is the number of clusters with internal
velocity dispersion greater than $\sigma_{cut}$, $N_{gal}$ is the
total number of galaxies in those clusters, $f_{cl}$ is the percentage
of galaxies in clusters, and $n_{cl}$ is the number density of
clusters. }

\end{table*}

\clearpage

\begin{table*}
\begin{tabular}{cccc}
Simulation & $\overline{\sigma}_{cl} (km/s)$ & $\overline{N}_{gal}$ &
$\sigma_{12}(1)$\\
\tableline
CDM1   & $550 \pm 49$ & $20 \pm 16$ & $745 \pm 114$\\
CDM1.5 & $562 \pm 45$ & $17 \pm 9$ & $605 \pm 38$ \\
CHDM1  & $552 \pm 46$ & $21 \pm 10$ & $652 \pm 37$\\
CHDM2  & $533 \pm 23$ & $25 \pm 22$ & $508 \pm 41$\\
\tableline

\end{tabular}

\caption{Properties of clusters with $\sigma_{cut} \ge 500$
km/s. The second column shows $\overline{\sigma}_{cl}$, the mean
internal line-of-sight velocity dispersion of the clusters, with the
standard deviation shown as the error. Column 3 shows the mean and
standard deviation of the number of galaxies per cluster. Column 4 is
$\sigma_{12}(1)$ after the clusters were removed. }

\end{table*}

\begin{table*}
\begin{tabular}{ccccc}
Simulation & $\sigma_{DP} (F=1)$ & $\sigma_{DP}$ (no streaming, F=0) &
$\sigma_{true}$ \\
\tableline
CDM1 & $1024 \pm 145$ & $878 \pm 133$ & $862 \pm 33$ \\
CDM1.5 & $764 \pm 71$ & $658 \pm 62$ & $677 \pm 41$ \\ 
CHDM1 & $736 \pm 43$ & $607 \pm 40$ & $656 \pm 52$ \\
CHDM2 & $537 \pm 40$ & $444 \pm 35$ & $543 \pm 72$ \\

\end{tabular}

\caption{Comparison of $\sigma_{12}$ computed using the method of
Davis and Peebles with $\sigma_{true}$, computed by fitting directly
to the pairwise peculiar velocity distribution. }

\end{table*}

\clearpage

\begin{figure} 
\caption{Contours show $\xi(r_{p}, \pi)$ averaged over
six views and smoothed. Contour spacings are $\Delta \xi = 0.1$ for
$\xi<1$ and $\Delta \log \xi = 0.1$ for $\xi > 1$, with the solid
contour indicating $\xi=1$.  }
\label{fig:xicont}
\end{figure}

\begin{figure} 
\caption{The velocity dispersion for the 10.4 sr sky-catalogs, averaged over
six views, with the standard deviation shown as error bars. The text
markers CfA2N, CfA2S, and SSRS2 show the values from Marzke et
al. (1995), for the CfA2/SSRS2 survey. They are shown offset to
$r_{p}=1$ for visibility but actually were calculated for exactly the
same bin in $r_{p}$ as our result.  }
\label{fig:sigma}
\end{figure}

\begin{figure}
\caption{The velocity dispersion for the 10.4 sr sky-catalogs with clusters
removed, averaged over six views, with standard deviations shown as
error bars. The text markers CfA2N, CfA2S, and SSRS2 show results from
Marzke et al. (1995) for CfA2/SSRS2 with Abell clusters with richness
$R \geq 1$ removed. See comment with Figure~\ref{fig:sigma} on the
horizontal offset.  }
\label{fig:sigmanoclusters} 
\end{figure} 

\begin{figure} 
\caption{The velocity dispersion $\sigma_{12}(1)$, with clusters
removed at different internal cluster velocity dispersion thresholds
($\sigma_{cut}$). These values are for one mock catalog (view a). }
\label{fig:sigc_vs_sig12} 
\end{figure}

\begin{figure} 
\caption{The change in velocity dispersion at $r=1h^{-1}$ Mpc
($\Delta\sigma_{12}(1) = \sigma_{12}(1) - \sigma_{12}(1)_{clusters
\;cut}$) as clusters are removed at different thresholds, as a function
of the number of clusters cut ($N_{cl}$).  These results are for one
mock catalog (view a). }
\label{fig:ncut_vs_dsig12} 
\end{figure}

\begin{figure} 
\caption{The first moment of the pairwise peculiar velocity distribution, or
mean streaming $\overline{v_{12}}(r)$. The dotted line is the BBGKY
model (equation~\ref{eq:infall}) with $F=1$, the dashed line is the
model with F=0.5, and the smooth solid line is the Hubble flow ($H_0 =
50$ km/s/Mpc). Results are shown for all six views in each
simulation. }
\label{fig:v12} 
\end{figure}  

\begin{figure} 
\caption{The velocity dispersion for the simulations with F=0 used in the
streaming model (equation~\ref{eq:infall}), averaged over six views,
with standard deviations shown as error bars. The text markers CfA2N,
CfA2S, and SSRS2 show results with F=0 from Marzke et al. (1995) for
CfA2/SSRS. See comment with Figure~\ref{fig:sigma} on the horizontal
offset.  }
\label{fig:signoinfall} 
\end{figure}

\begin{figure} 
\caption{The second moment of the pairwise peculiar velocity
distribution $f(w_3 \mid r)$, obtained by fitting the exponential
model directly to $f(w_3 \mid r)$.  }
\label{fig:truesigma} 
\end{figure}


\clearpage

\begin{references}

\reference{} Bartlett, J.G. \& Blanchard, A. 1995, preprint

\reference{} da Costa, L.N., Pellegrini, P.S., Davis, M., Meiksin, A.,
Sargent, W.L., \& Tonry, J.L. 1991, \apjs, 75, 935

\reference{} Davis, M., 1995, personal communication.

\reference{} Davis, M., Efstathiou, G., Frenk, C. S., White, S. D. M.
1985, \apj, 292, 371

\reference{} Davis, M. \& Huchra, J. 1982, \apj, 254, 437

\reference{} Davis, M. \& Peebles, P.J.E. 1977, \apjs, 34, 425

\reference{} Davis, M. \& Peebles, P.J.E. 1983, \apj, 267, 465 (DP83)

\reference{} Davis, M. 1988, in Cosmology and Particle Physics, ed.
L.Z. Fang \& A. Zee (New York: Gordon and Breach), 65.

\reference{} Evrard, A.E., Summers, F.J., \& Davis, M. 1994 \apj, 422, 11

\reference{} Fisher, K.B., Davis, M., Strauss, M.A., Yahil, A. \&
Huchra,~J. P. 1994a, \mnras, 266, 50
  
\reference{} Fisher, K. B., Davis, M., Strauss, M.A., Yahil, A. \&
Huchra,~J.P. 1994b, \mnras, 267, 927

\reference{} Frenk, C.S., White, S.D.M., Efstathiou, G., \& Davis, M.
1990, \apj, 351, 10

\reference{} Gelb, J.M., \& Bertschinger, E. 1994, \apj 436, 491

\reference{} Guzzo, L., Fisher, K.B., Strauss, M.S., Giovanelli, R. \&
Haynes,~M.P. 1995, preprint

\reference{} Hernquist, L., \& Katz, N. 1989, \apjs, 70, 419 

\reference{} Huchra, J. P., Davis, M., Latham, D.W. \& Tonry, J. 1983,
\apjs, 52, 89

\reference{} Katz, N., \& White, S.D.M. 1993, \apj, 412, 455

\reference{} Klypin, A., Holtzman, J., Primack, J. \& Regos, E. 1993,
\apj, 416, 1
 
\reference{} Klypin, A., Nolthenius, R., Primack, J. 1996, \apj, in
press (KNP)

\reference{} Liddle, A., Lyth, D.H., Schaefer, R.K., Shafi, Q., \&
Viana, P.T.P.  1995, preprint
  
\reference{} Marzke, R. O., Geller, M.J., daCosta, L.N. \& Huchra, J.
P. 1995, \aj, 110, 477

\reference{} Mo, H. J., Jing, Y.P. \& Borner, G. 1993, \mnras, 264, 825

\reference{} Nolthenius, R., Klypin, A., \& Primack, J. 1996, \apj,
submitted (NKP)

\reference{} Peebles, P.J.E. 1976, \apss, 45, 3 

\reference{} Peebles, P.J.E. 1980, The Large Scale Structure of the
Universe (Princeton: Princeton University Press)

\reference{} Primack, J.R., Holtzman, J., Klypin, A., \& Caldwell, D.O.
1995, Phys. Rev. Lett., 74, 2160

\reference{} Smoot, G.F. et al. 1992, \apj, 396, L1

\reference{} Somerville, R.S., Davis, M., \& Primack, J.R. 1996,
preprint (SDP)

\reference{} Summers, F.J., Davis, M., \& Evrard, A. 1995, \apj, 454, 1

\reference{} Zurek, W., Quinn, P.J., Salmon, T.K., \& Warren, M.S.
1994, \apj, 431, 559

\end{references}
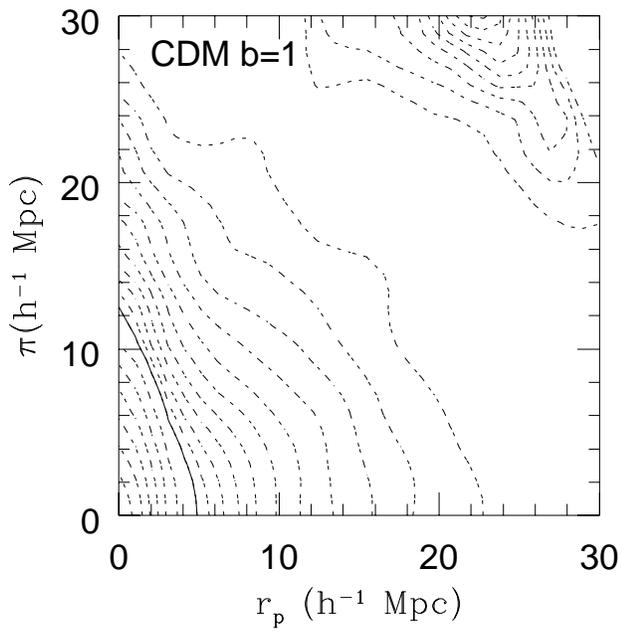
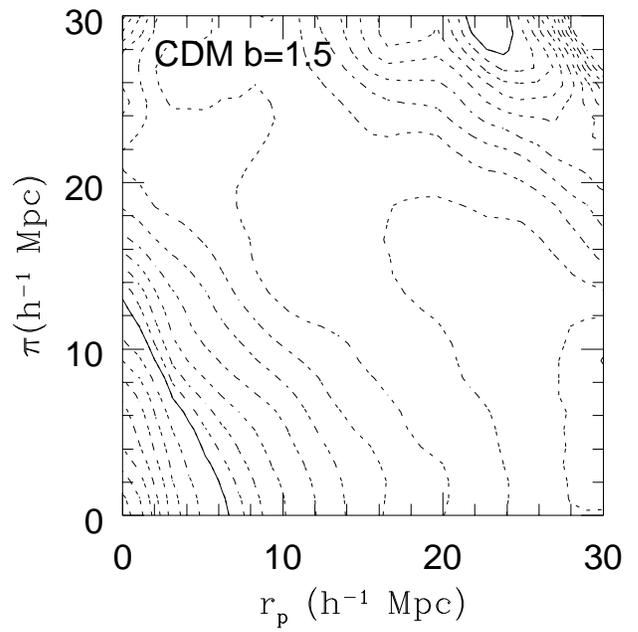
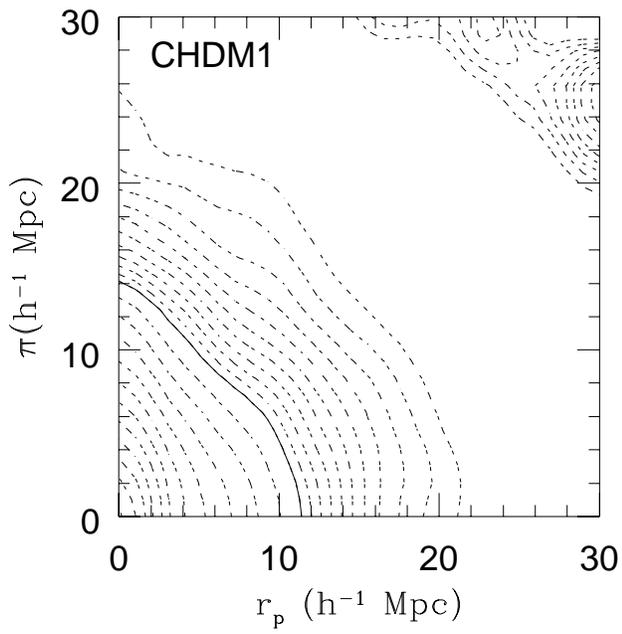
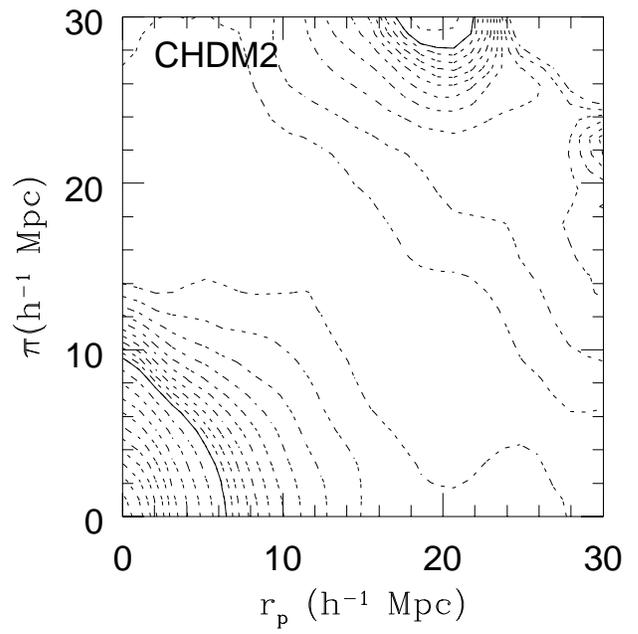
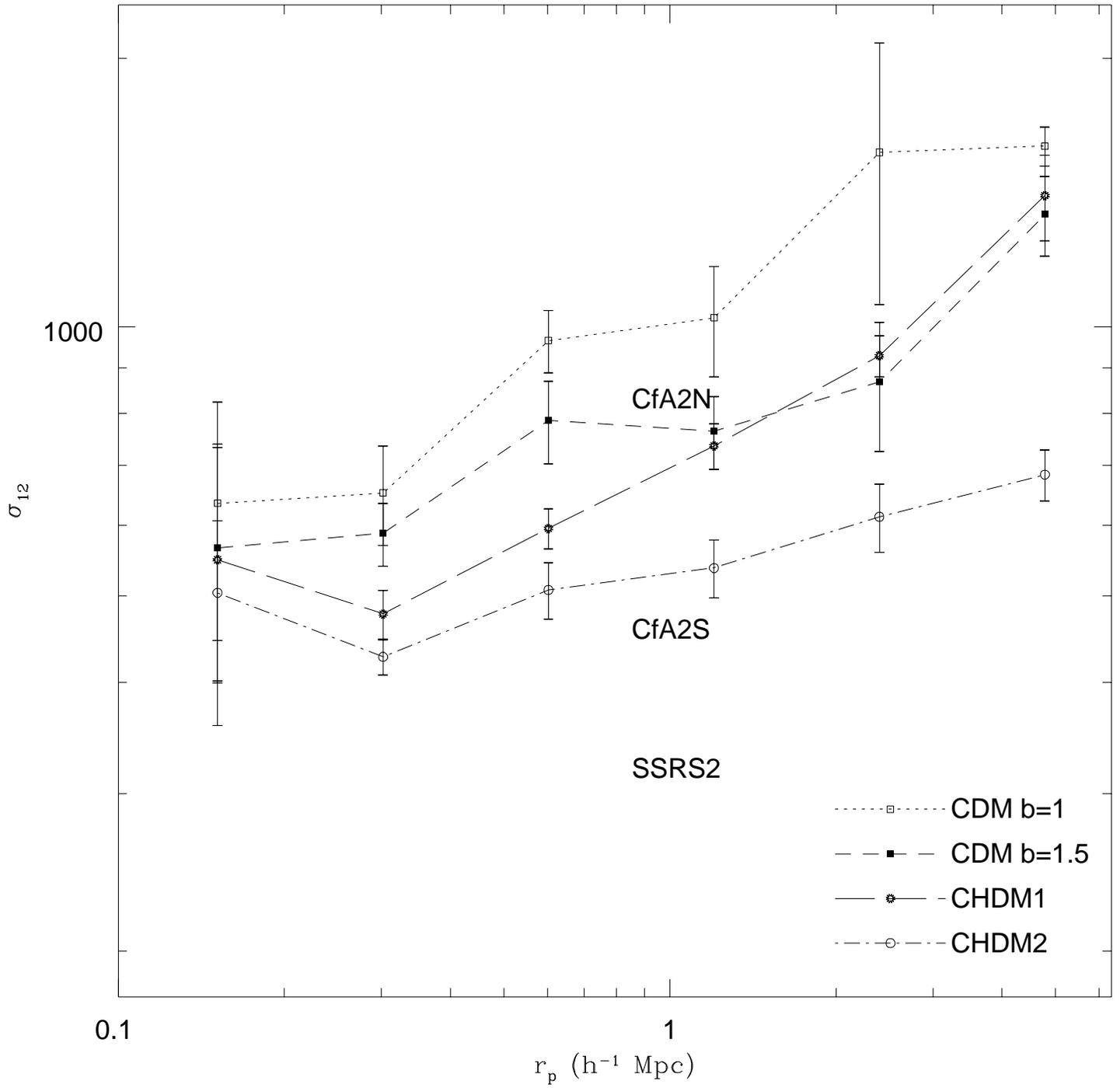
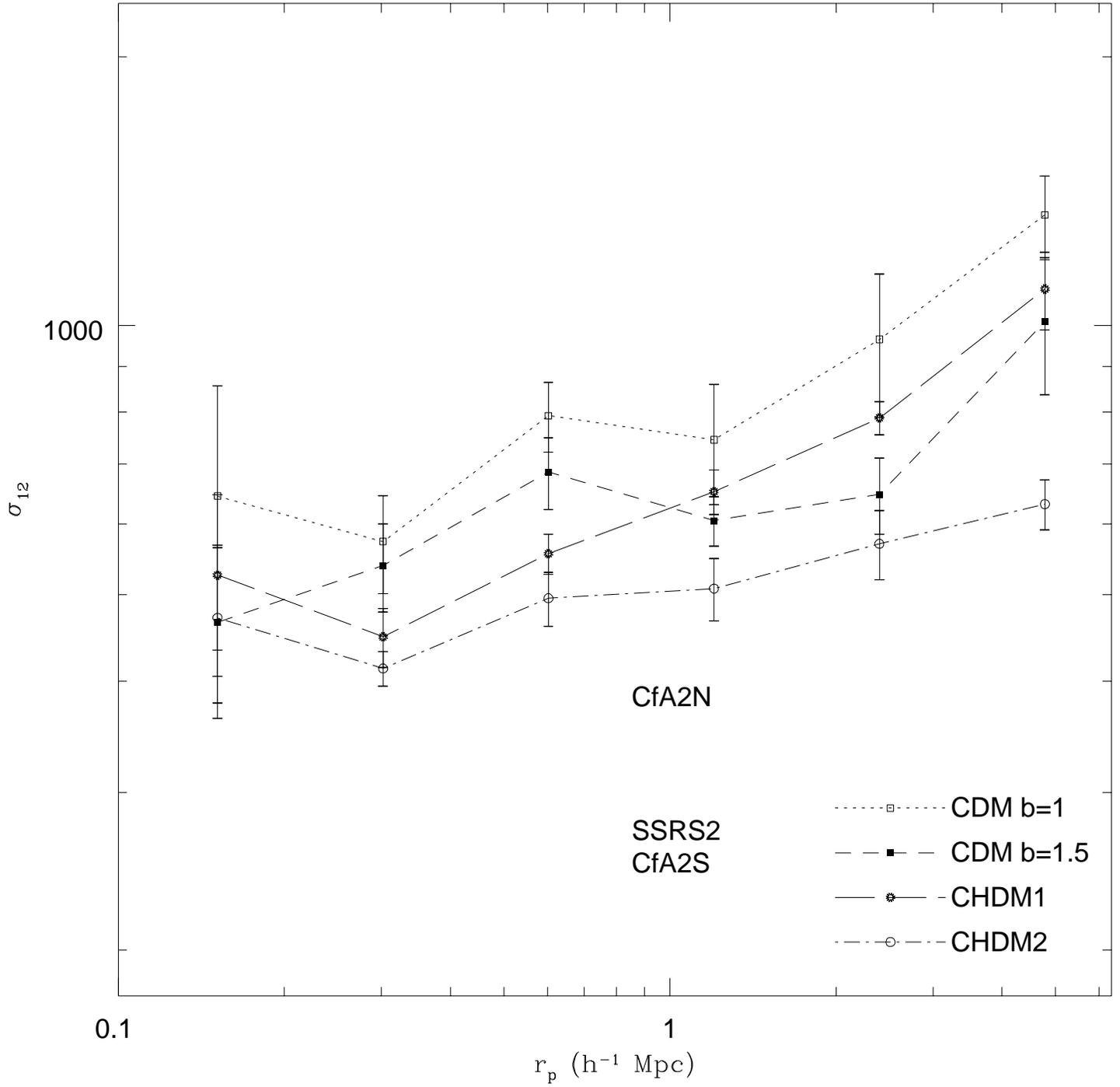
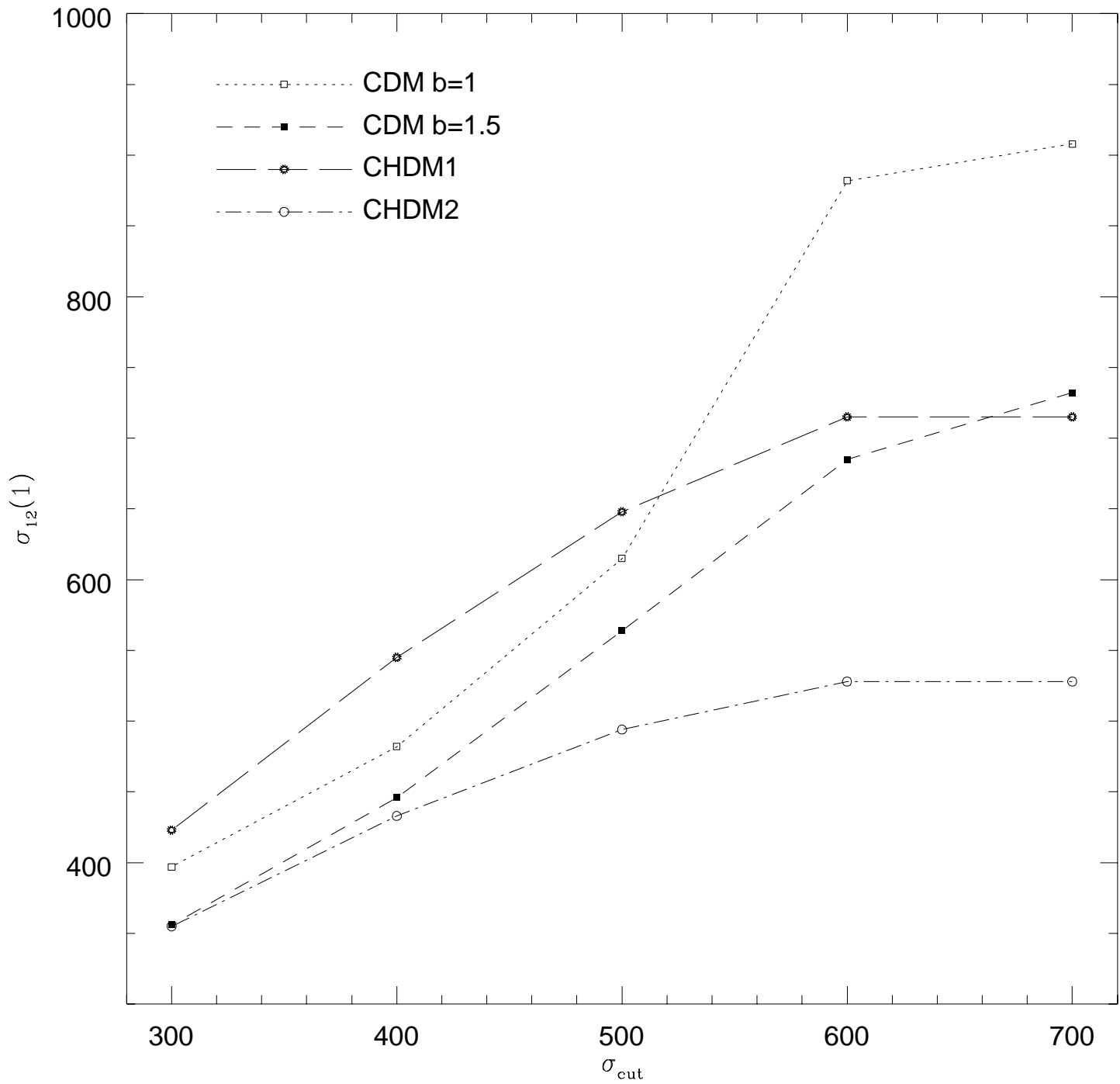
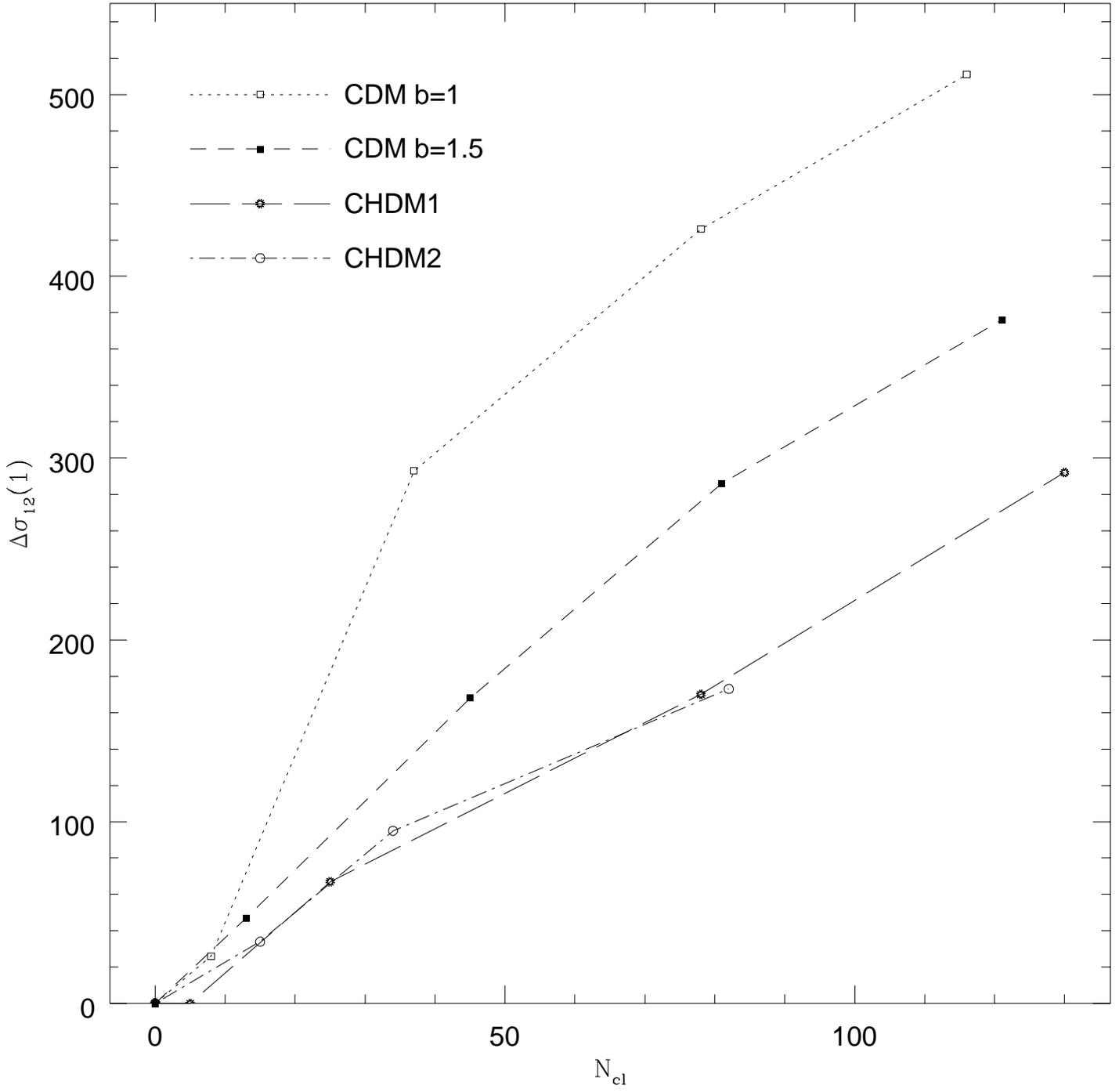
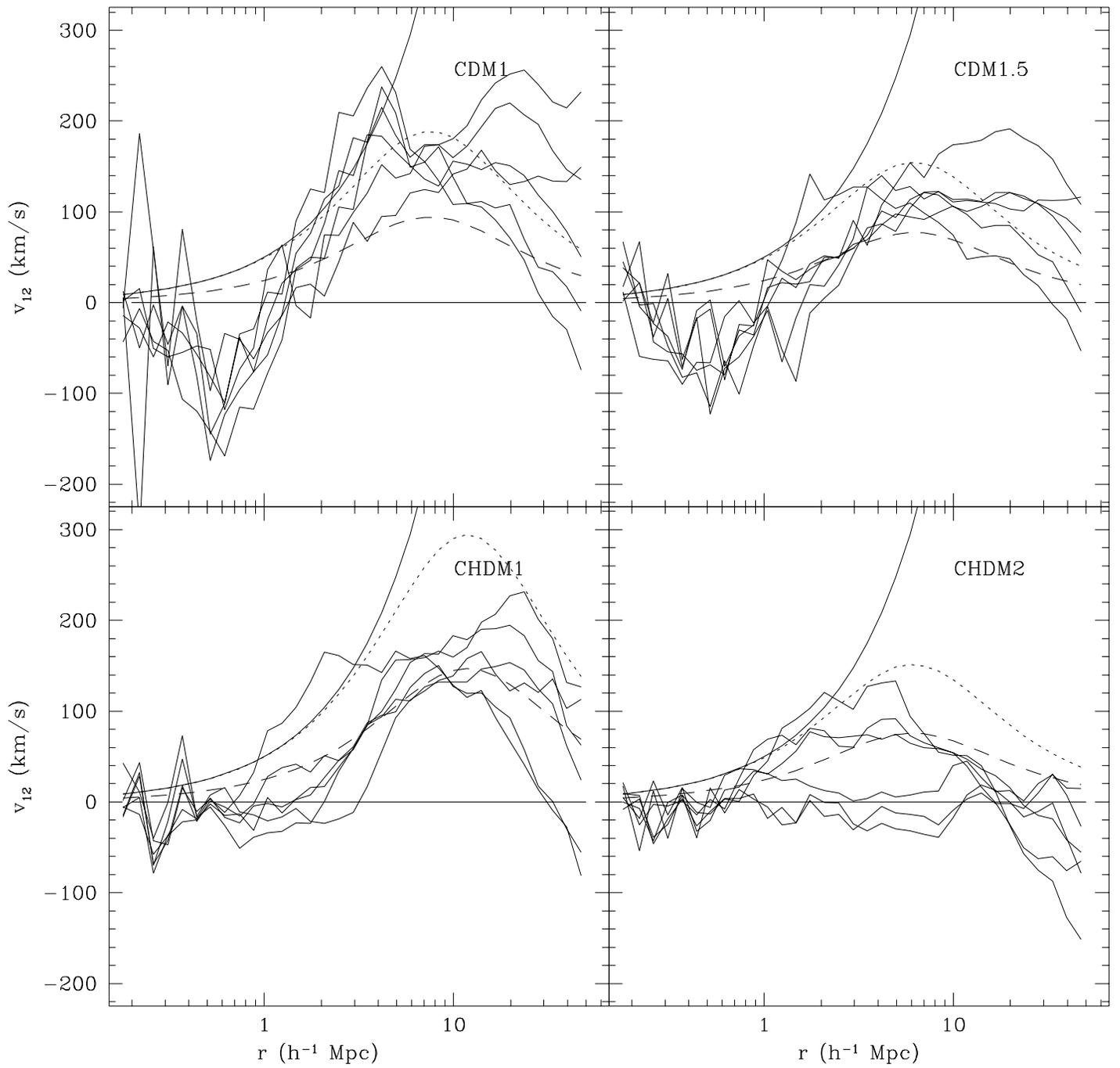
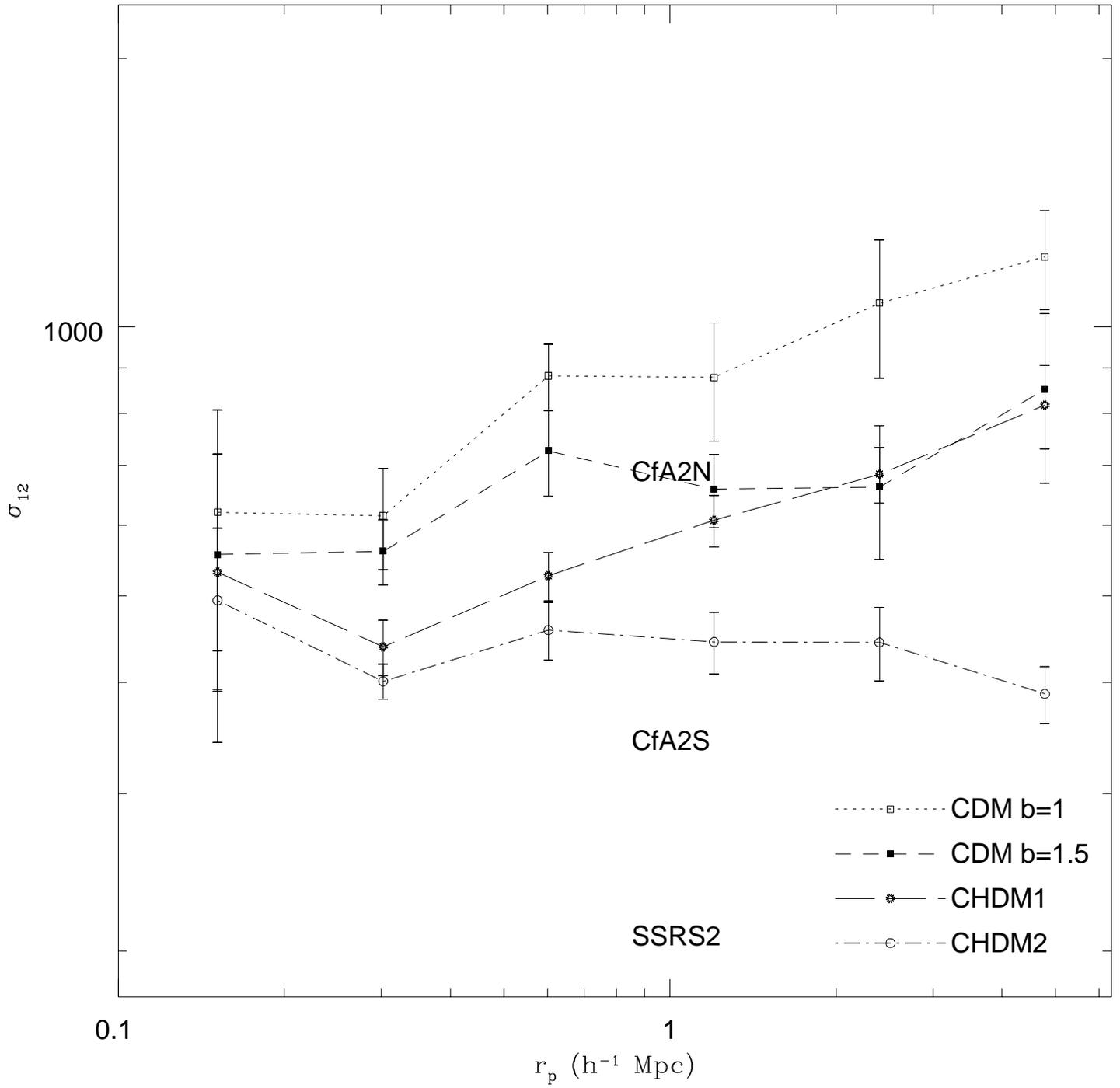
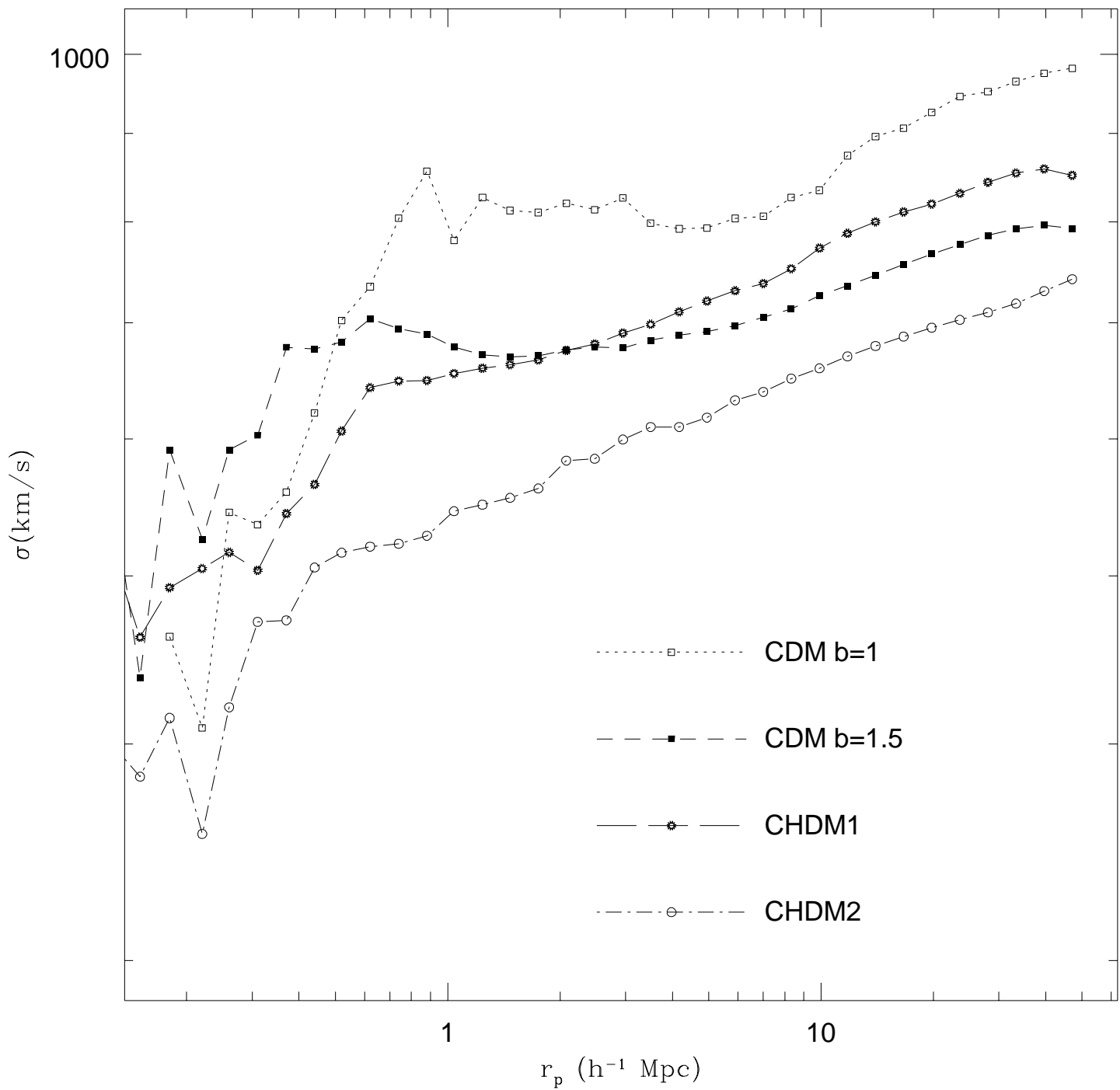
\end{document}